\documentclass[12pt]{emulateapj}
\usepackage{apjfonts}

\newcommand{\pfrac}[2]{\left(\frac{#1}{#2}\right)}
\def\be{\begin{equation}}
\def\ee{\end{equation}}
\def\bea{\begin{eqnarray}}
\def\eea{\end{eqnarray}}
\def\f{\frac}

\def\v{\varepsilon}
\def\e{\epsilon}
\def\bt{s} 
\def\hmu{\tilde{\mu}}

  

\shorttitle{The bulk Lorentz factors of GRBs} \shortauthors{Zhao,
Li \& Bai}

\begin{document}

\title{The bulk Lorentz factors of Fermi-LAT GRBs}

\author{Xiao-Hong Zhao\altaffilmark{1,4}, Zhuo Li\altaffilmark{2,3}, Jin-Ming Bai\altaffilmark{1,4}}

\altaffiltext{1}{National Astronomical Observatories/Yunnan
Observatory, Chinese Academy of Sciences, P.O. Box 110, 650011
Kunming, China; zhaoxh@ynao.ac.cn}
 \altaffiltext{2}{Department of Astronomy, Peking University, Beijing 100871,
 China}
 \altaffiltext{3}{Kavli Institute for Astronomy and Astrophysics, Peking University, Beijing 100871,
 China}
 \altaffiltext{4}{Key Laboratory for the Structure and Evolution of Celestial Bodies,
Chinese Academy of Sciences, P.O. Box 110, 650011 Kunming, China }

\begin{abstract}
The Lorentz factor (LF) of gamma-ray burst (GRB) ejecta may be
constrained by observations of high-energy (HE) spectral
attenuation. The recent Fermi-LAT observations of prompt GeV
emission from several bright GRBs have leaded to conclusions of
unexpectedly large LFs, $\Gamma>10^3$. Here we revisit this problem
with two main concerns. (1) With one-zone assumption where all
photons are assumed to be generated in the same region (radius) and
time, we {\em self-consistently} calculate the $\gamma\gamma$
optical depth by adopting a target photon spectrum with HE cutoff.
We find that this might be important when the GRB LF is below a few
hundreds. (2) Recent Fermi-LAT observations suggest that the bulk
MeV-range and HE ($\ga100$~MeV) emission may arise from different
regions. We then consider a two-zone case where HE emission is
generated in much larger radii than that of the MeV-range emission.
We find that the HE emission may be mainly attenuated by MeV-range
emission and that the attenuated HE spectrum does not show an
exponential spectral cutoff but a slight steepening. This suggests
that there may be no abrupt cutoff due to $\gamma\gamma$ attenuation
if relaxing the one-zone assumption. By studying the spectra of
three bright Fermi-LAT GRBs 080916C, 090510 and 090902B, we show
that a bulk LF of\textbf{ $\Gamma\sim600$ }can be consistent with
observations in the two-zone case. Even lower LFs can be obtained in
the multi-zone case.
\end{abstract}

\keywords{Gamma ray burst: general}

\section{Introduction}
Relativistic expansion is a key property of gamma-ray bursts (GRBs),
and has been confirmed by measurements of radio afterglow sizes, for
examples, the indirect estimation by radio scintillation in GRB
970508 \citep{wkf98} and direct imaging of nearby GRB 030329
\citep{taylor04}. These observations revealed mildly relativistic
GRB ejecta, $\Gamma\sim$a few, in the radio afterglow phase.
However, it is well believed that GRB ejecta are ultra-relativistic
in the beginning-- this is required to solve the so-called
"compactness problem" \citep[e.g.,][]{piran99}. The compact GRB
source, suggested by the rapid variabilities in MeV light curves,
and the huge luminosity suggest hot, optically thick GRB sources,
which is in confliction with the nonthermal and hard GRB spectra.
Relativistic expansion of the emission region is introduced to solve
this problem. In order for the $\sim100$~MeV photons, as detected by
EGRET in several GRBs, to escape from the emission region, avoiding
$\gamma\gamma$ attenuation, the bulk Lorentz factor (LF) of the
emission region is required to be extremely large, $\Gamma\ga10^2$
\citep[e.g.,][]{ls01,krolik91,fenimore93,wl95,bh97}. Recently, the
powerful Fermi satellite reveals in much more detail the high-energy
(HE) emission from GRBs. Several bright GRBs are reported to show
time-integrated spectra extending up to GeV or even tens GeV,
without any signs of spectral cutoff. Assuming the $\gamma\gamma$
optical depth for these HE photons are below unity, these
observations have leaded to even larger bulk LFs, $\Gamma>10^3$
\citep{a09a,a09b,a09c}. This is putting the theoretical problem of
relativistic jet formation to extremes.

In the previous constraints two assumptions are usually taken.
First, all photons, from low to high energy, are produced in the
same region and the same time. This "one-zone" assumption is not
solid, as Fermi observations actually revealed that: the onset of
HE emission is delayed relative to MeV emission
\citep[e.g.,][]{a09a,a09b,a09c}; the HE emission lasts longer than
MeV emission \citep[e.g.,][]{a09a,a09b,a09c}; the bulk emission
shifts toward later time as the photon energy increases
\citep{a09a} and the shift is longer than the variability times in
MeV light curves, as pointed out by \citep{li10}; some GRBs
obviously show distinct HE components with different temporal
behaviors \citep{a09b,a09c}. All these features may imply that
different energy photons are produced in different regions.

In particular, the bulk $>100$~MeV emission in GRB 080916C shows
$\sim1$~s shifting relative to MeV emission, which is much longer
than the MeV variability time, $<100$~ms as revealed by INTEGRAL
\citep{greiner09}, strongly implying that $>100$~MeV emission is
produced in a region of much larger radii than MeV emission's
\citep{li10}. As pointed out by \cite{lw08}, within the framework of
internal shock model, the internal collisions at small radii, which
would produce the prompt MeV emission, are expected to lead to
"residual" collisions at much larger radii, which would produce
low-frequency emission. The electrons accelerated by residual
collisions at larger radii inverse-Compton scattering the MeV
photons and/or double scattering the low-frequency photons could
produce HE emission \citep{li10,zhao10}. In this case, MeV and HE
photons are produced in different regions. In the comoving frame of
HE emission region, the MeV photons would be collimated other than
isotropic, thus the $\gamma\gamma$ absorption is angular dependent.

Second, the target photon spectrum is assumed to be extending to
infinity. As pointed out by \cite{li10}, the calculation of
$\gamma\gamma$ optical depth taking such a target photon filed is
obviously not self-consistent, because the HE spectral end should
be cut off due to absorption considered in the calculation.

In this paper, we revisit the problem of GRB LF constraint by
modifying the above mentioned two assumptions. We consider in \S2
a one-zone case where the $\gamma\gamma$ optical depth is
calculated {\em self-consistently} by assuming a truncated target
spectrum, then we consider in \S3 a simple two-zone case with
anisotropic effect on $\gamma\gamma$ optical depth taken into
account. In \S4 we studied the spectra of the three bright
Fermi-LAT GRBs and constrain their LFs. \S5 is discussion and
conclusions. In the following we assume the concordance universe
model with $(\Omega_m,\Omega_\Lambda)=(0.27,0.73)$ and $H_0=71\rm
km\,s^{-1}Mpc^{-1}$.

\section{One-zone case}
Consider a GRB ejecta with bulk LF $\Gamma$ and radius $R$. Assume
the photons in the comoving frame of the ejecta is isotropic, with
photon number density per photon energy $dn'/d\e'$. Hereafter,
unless specified otherwise, quantities with prime denote the
comoving frame, and non-primed ones denote the frame of observer on
the Earth.

In the (comoving-frame) dynamical time $R/\Gamma c$, a photon
travels a path of $R/\Gamma$. For a photon of energy
$\v'=\v(1+z)/\Gamma$ (with $z$ the GRB redshift), the optical
depth due to $\gamma\gamma$ collisions during a dynamical time is
given by \citep{Gould67}
\begin{equation}\label{eq:tau}
  \tau(\varepsilon^\prime)=\f{R}{2\Gamma}\int_{m_e^2c^4/\varepsilon^\prime}^{\v'_{\max}}d\epsilon^\prime\f{dn^\prime}{d\epsilon^\prime}
\int^1_{-1}d\hmu^\prime(1-\hmu^\prime)\sigma(E),
\end{equation}
where $\hmu'=\cos\Theta'$ and $\Theta'$ is the angle between the
colliding photon pair. The cross section is given by
\begin{equation}
  \sigma(E)=\f{3\sigma_T}{16}(1-\beta_e^2)\left[(3-\beta_e^4)\ln\f{1+\beta_e}{1-\beta_e}-2\beta_e(2-\beta_e^2)\right]
\end{equation}
where $\beta_e=\sqrt{1-(m_ec^2/E)^2}$ and
$E=\sqrt{\varepsilon^\prime\epsilon^\prime(1-\hmu^\prime)/2}$ are
the velocity and energy, respectively, of the generated electron
in the center of momentum frame of the collision. The radius $R$
of the emission region can be related to the angular spreading
time $\delta t_{\rm ang}$, due to geometry effect, by
$R=2\Gamma^2c\delta t_{\rm ang}/(1+z)$. As the angular spreading
time is related to the observed variability time $\delta t$ by
$\delta t_{\rm ang}=\delta t$, we have
\begin{equation}\label{eq:r}
  R=2\Gamma^2c\frac{\delta t}{1+z}.
\end{equation}
For a GRB with the observed photon number per unit time per unit
photon energy per unit detector area, denoted by $N(\e)$, the
photon number density per unit photon energy in the comoving frame
can be given by
\begin{equation}\label{eq:densitydistr}
  \frac{dn'}{d\e'}=\pfrac{d_L}{R}^2\frac{N(\e)}{c(1+z)^2},
\end{equation}
where $d_L$ is the GRB luminosity distance, and
$\e=\Gamma\e'/(1+z)$.

It is important to note a difference from the previous works. In eq.
(\ref{eq:tau}) we did not take the upper limit of the integration to
be infinity but a certain photon energy $\v'_{\max}$, because the HE
tail is expected to be cut off due to $\gamma\gamma$ absorption. The
cutoff energy is just where $\tau(\v'_{\max})=1$ happens. To
self-consistently solve out the cutoff energy
$\v_{\max}=\Gamma\v'_{\max}/(1+z)$ for given $\Gamma$, we need to
take the upper limit of the integration to be $\v'_{\max}$, and
solve $\tau(\v'_{\max})=1$ using eqs.
(\ref{eq:tau}-\ref{eq:densitydistr}) and observed GRB spectrum
$N(\e)$.

It is well known that the GRB spectrum can be fit by the Band
function \citep{band93}
\begin{equation}
 N(\epsilon) = \left\{\begin{array}{cc}
A(\f{\epsilon}{100{\rm keV}})^{ \alpha} \exp
\big[-\frac{\epsilon(2+\alpha)} {\epsilon_p}\big] \
& \epsilon < \epsilon_c \\
A\big[\f{( \alpha-\beta) \epsilon_p}{(2+\alpha) 100{\rm keV}}
\big]^{( \alpha-\beta)}  \exp ( \beta- \alpha)
(\f{\epsilon}{100{\rm keV}})^{\beta} & \epsilon >
\epsilon_c\end{array}, \right.
\end{equation}
where $\epsilon_c=\epsilon_p(\alpha-\beta)/(2+\alpha)$, and $A$,
$\alpha$, $\beta$ and $\epsilon_p$ are the normalized coefficient,
low-energy slope, HE slope and the $\nu F_\nu$ peak energy,
respectively. In some Fermi-LAT GRBs an extra spectral component
beyond the Band-function is claimed to exist, especially in HE end
\citep{a09b,a09c}. This extra component can be described as a
power law,
\begin{equation}
  N(\e)=A_{\rm PL}\pfrac{\e}{1\rm GeV}^{\beta_{\rm PL}},
\end{equation}
with $A_{\rm PL}$ the normalization at 1~GeV and $\beta_{\rm PL}$
the spectral index.

It is helpful to solve out the $\Gamma-\v_{\max}$ relation with
some approximations first. Typically the HE, $\ga100$~MeV, photons
mainly interact with photons above the peak energy. Let us
approximate the target photon distribution as a single power law
$N(\e)=N_0\e^{-\bt}$ in the following analytical derivation.

In eq. (\ref{eq:tau}), usually the upper limit of the first
integral is taken to be $\infty$. This is valid for
$\varepsilon_{\max}\gg\Gamma^2m_e^2c^4/[\varepsilon_{\max}(1+z)^2$]
and the spectrum slope $\bt>1$. In this case, using
$\delta$-approximation for the cross section at target photon
energy above the threshold, $\sigma\approx(3/16)\sigma_T$,
$\tau(\v_{\max})=1$ can be solved to give $\Gamma$ as function of
$\v_{\max}$,
 \bea
\Gamma\propto\v_{\max}^{\frac{1+s}{2(s-1)}}.
 \eea
However, when
$\varepsilon_{\max}\gtrsim\Gamma^2m_e^2c^4/[\varepsilon_{\max}(1+z)^2]$,
i.e., the energy of annihilated photons is compared with that of
target photons, the upper limit cannot be taken as $\infty$ any
more. In the case, $\Gamma$ is given by \citep{li10}
 \bea
\Gamma\approx\f{\varepsilon_{\max}}{m_ec^2}(1+z).
 \eea

\begin{figure}[h]
\includegraphics[width=\columnwidth]{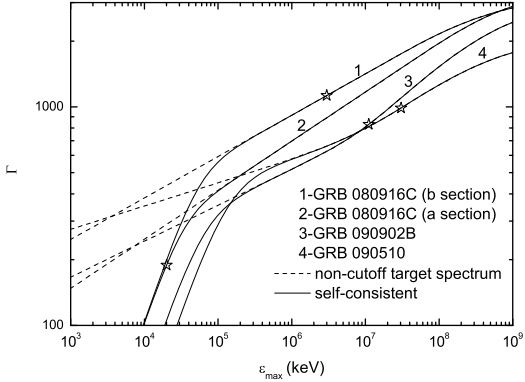}
\caption{The relation between the observed maximum photon energy
and the lower limit to the bulk LF in the one-zone case for the
three bright GRBs. The adopted parameters of the GRBs are shown in
Table \ref{tab1}. As marked in the plot, the dash lines correspond
to results using target photon without spectral cutoff, while the
solid lines correspond to our {\em self-consistent} calculations
using truncated target photon spectra. The stars denote the
observed highest energy of photons in the relevant time
intervals.}\label{fig:1zone}
\end{figure}

Next we carry numerical calculation to solve out
$\tau(\v_{\max})=1$. For the observations, we take the three
bright Fermi-LAT GRBs 080916C, 090510 and 090902B, and consider
the same time intervals in the GRBs where the LFs have been
constrained by \cite{a09a,a09b,a09c}, as well as section a in GRB
080916C. The properties of spectra and flux for these GRBs are
shown in Table \ref{tab1}. The calculated results are given in Fig
\ref{fig:1zone}, where we compare the results of self-consistent
calculation and previous method using a target photon spectrum
without HE cutoff. We see that the results deviate each other for
$\v_{\max}\la100$~MeV or $\Gamma\la$a few hundreds. In the case of
section a in GRB 080916C, where the maximum observed photon energy
is lower (see Fig \ref{fig:1zone}), the Lorentz factor limit with
the self-consistent calculation is much smaller than that with the
previous method. Thus to be self consistent, the upper bound of
the integration in eq (\ref{eq:tau}) should be carefully taken as
the maximum photon energy in this case. We also note that the LF
constraints using upper limit of infinity are still valid for
those time segments that have been used by \cite{a09a,a09b,a09c}.

\begin{deluxetable*}{lccccccccccc}
\tabletypesize{\scriptsize} \tablecaption{The parameters of three
bright LAT-GRBs} \tablewidth{0pt}
 \tablehead{\colhead{GRB name}&\colhead{Time interval}&\colhead{$\epsilon$$_{p}$}&\colhead{$\alpha$}&\colhead{$\beta$}&\colhead{$A$}&
$\beta_{\rm PL}$ & $A_{\rm PL}$  &$z$& $\delta t$   &   $\v_{\rm highest}$ \\
 & (s) & (keV) &  &  & (cm$^{-2}$s$^{-1}$keV$^{-1}$) & & (cm$^{-2}$s$^{-1}$keV$^{-1}$)&  & (ms) & (GeV)  }
 \startdata
  GRB 080916C-a &   0.004-3.58  &  440  & $-0.58$ & $-2.63$ & 0.055 & --  & -- &4.35 & 100$^a$ & 0.02$^b$  \\
  GRB 080916C-b &   3.58-7.68  &  1170  & $-1.02$ & $-2.21$ & 0.035 & --  & -- &4.35 & 100$^a$ & 3  \\
  GRB 090510  & 0.8-0.9   &  1894  & $-0.86$ & $-3.09$ & 0.028 &  $-1.54$ & 6.439$\times$10$^{-9}$&0.903  & 12 &30.5 \\
  GRB 090902B & 9.6-13   &  821  & $-0.26$  & $-5.0$  &  0.082 & $-1.98$ & 4.3$\times$10$^{-10~c}$ & 1.822& 53  &11.2
\enddata
\tablecomments{References: a: \cite{greiner09}; b: the spectral
flux of this section at $>20$~MeV is only upper limit, as shown in
the supporting material of \cite{a09a}; c: private communication
with Francesco de Palma; and the other parameters are taken from
\cite{a09a,a09b,a09c}. }\label{tab1}
\end{deluxetable*}

\section{Two-zone case}
As discussed in the introduction, the Fermi-LAT observations hint
that there may be different emission regions of different radii in
GRB prompt emission. As the HE delay of onset and the shifting of
the bulk HE emission are in seconds scale, whereas the MeV-range
variability times, reflecting the dynamical time of the MeV
emission region, are in tens of ms scale, the MeV emission regions
have much smaller, by orders of magnitude, size (radius) than that
of HE emission regions.

Consider that the ejecta expand to radius $R$ where HE emission is
being produced. The photons that are emitted in much smaller radii
and just arrive at radius $R$ should be produced by those ejecta
released from the central engine with a time delay of
$t_d=R(1+z)/2\Gamma^2c$. If $t_d\gg\delta t$ then $t_d$ is also
the observed delayed time scale of the HE emission. Thus, once we
observed a time delay $t_d(\gg\delta t)$ for HE emission relative
to MeV emission, the HE emission size is implied to be
\begin{equation}\label{eq:rdelay}
  R=2\Gamma^2c\frac{t_d}{1+z}.
\end{equation}

As the MeV emission comes from inner regions with smaller radii,
$R_{\rm MeV}=2\Gamma_{\rm MeV}^2c\delta t/(1+z)\ll R$, the MeV
photons in the comoving frame of the HE emission region are
beamed. Here we also denote the LF of MeV emission region as
$\Gamma_{\rm MeV}$ since it may be different from the one of the
HE emission region, $\Gamma$, in the framework of internal shock
model, and the difference could be small, $\Gamma_{\rm
MeV}\sim\Gamma$.

\begin{figure}[h]
\includegraphics[width=\columnwidth]{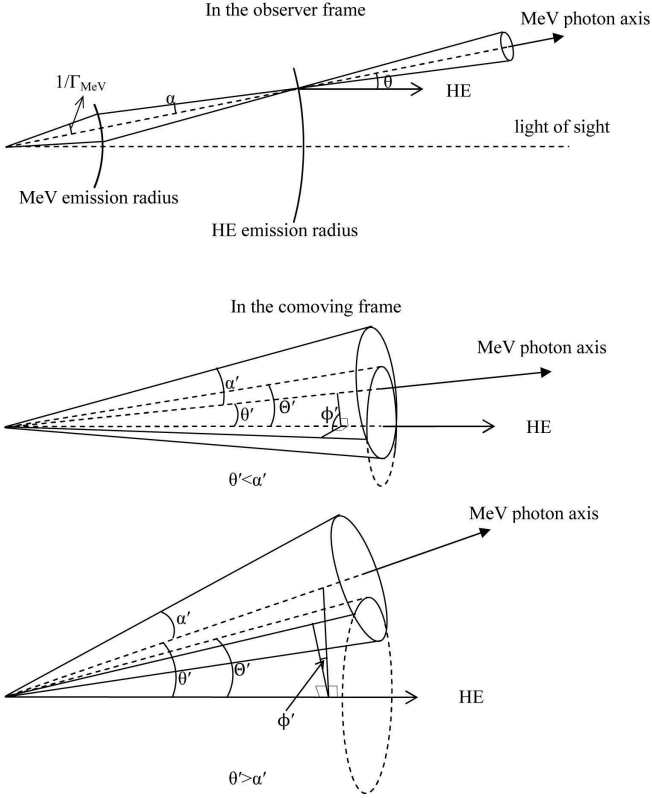}
\caption{The schematic diagram of the geometry when the a HE
photon produced at large radius $R$ is being illuminated by the
MeV photon beam from much smaller radii $R_{\rm MeV}$. The upper
panel is for the observer frame, while the bottom two panels for
the comoving frame of GeV emission region, with $\theta'<\alpha'$
and $\theta'>\alpha'$ respectively. }\label{fig:schematic}
\end{figure}

Consider the geometry plotted in Fig \ref{fig:schematic}. Due to
the relativistic beaming effect, the MeV emission beam that
illuminating a HE photon produced at $R$ can be approximated as a
"MeV photon cone" with half open angle of $\alpha=R_{\rm
MeV}/R\Gamma_{\rm MeV}$. Outside of the cone the MeV photon flux
can be neglected. In the comoving frame of HE emission region, the
solid angle is then $\Delta \Omega{^\prime}=2\pi
(1-\cos\alpha^\prime)$, with
$\cos\alpha^\prime=(\cos\alpha-\beta_\Gamma)/(1-\beta_\Gamma
\cos\alpha)$ and $\beta_\Gamma=\sqrt{\Gamma^2-1}/\Gamma$.

The optical depth is not only energy-dependent but also
angle-dependent. Consider a HE photon of $\varepsilon'$ travelling
with an angle $\theta'$ ($\mu'=\cos\theta'$) with respect to the
central axis of the target photon beam, then the optical depth
corresponding to the distance it travels in a dynamical time is
given by
\begin{eqnarray}\label{eq:tau_2zone_part_beam}
    \tau(\varepsilon^\prime,\mu^\prime)&=&\f{R}{\Gamma}\int_{
m^2c^4/\varepsilon^\prime}^{\infty}d\epsilon^\prime\int_{\{\Delta\Omega^\prime\}}
d\Omega^\prime\f{d^2n^\prime}{d\epsilon^\prime
d\Omega^\prime}(1-\hmu^\prime)\sigma(E) \nonumber\\
&=&\f{2R}{\Gamma\Delta \Omega^\prime}\int_{
m^2c^4/\varepsilon^\prime}^{\infty}d\epsilon^\prime\f{dn^\prime}{d\epsilon^\prime}\times\nonumber\\
&&\Bigg\{\begin{array}{ll}
\int^{\cos(\theta^\prime-\alpha^\prime)}_{\cos(\theta^\prime+\alpha^\prime)}d\hmu^\prime\phi^\prime
(1-\hmu^\prime)\sigma(E)   & \theta^\prime>\alpha^\prime \\
\\
\int^{1}_{\cos(\theta^\prime+\alpha^\prime)}d\hmu^\prime\phi^\prime
(1-\hmu^\prime)\sigma(E)  & \theta^\prime<\alpha^\prime.
\end{array}
\end{eqnarray}
Here $\hmu'=\cos\Theta'$ with $\Theta'$ the angle between HE
photon and the colliding target photon, $d^2n'/d\e'd\Omega'$ is
the energy distribution of target photons per unit solid angle,
with $d\Omega'=\sin\Theta' d\Theta' d\phi'$, and $\phi^\prime=\pi$
if $\theta'<\alpha'$ and $\hmu>\cos(\alpha'-\theta')$, otherwise
\begin{equation}
  \phi^\prime=\arccos\pfrac{\cos\alpha^\prime-\cos\Theta^\prime
\cos\theta^\prime}{\sin\Theta^\prime \sin\theta^\prime}.
\end{equation}
The comoving frame target photon density $dn'/d\e'$ is given by eq
(\ref{eq:densitydistr}) but with $\e=2\Gamma\e'/(1+z)$-- the
factor 2 appears for the highly beamed case of $\alpha'\ll1$.

In the extreme case when $R_{\rm MeV}/R\rightarrow0$ or
$\Gamma_{\rm MeV}/\Gamma\rightarrow+\infty$ the target photons are
totally beamed (hereafter other cases are called partly beamed) in
the comoving frame of the HE emission region, then
$\Delta\Omega^\prime\rightarrow 0$ and the optical depth reduces
to
\begin{eqnarray}\label{eq:tau_2zone}
    \tau(\varepsilon^\prime,\mu^\prime)=\f{R}{\Gamma}\int_{
2m^2c^4/\varepsilon^\prime(1-\mu^\prime)}^{\infty}d\epsilon^\prime\f{dn^\prime}{d\epsilon^\prime}(1-\mu^\prime)\sigma(E).
\end{eqnarray}

Consider an area element in the sphere emitting photons which lies
at an angle $\theta$ with respect to the line of sight, then in
its comoving frame a photon travelling along line of sight has an
angle with respect to the central axis of the target photon beam
of
\begin{equation}\label{eq:angletransfer}
  \mu'=\frac{\mu-\beta_\Gamma}{1-\beta_\Gamma\mu}.
\end{equation}
Due to Doppler effect, the photon energy in the comoving frame is
related to the observed photon energy as
\begin{equation}\label{eq:doppler}
  \varepsilon'=\Gamma(1-\beta_\Gamma\mu)\varepsilon(1+z).
\end{equation}
Denote $d^3P'/d\Omega'd\e'dS$ as the (comoving-frame) emitting
power per unit solid angle per unit photon energy by material in
per unit area of the sphere surface. This emission should be
modified by $\gamma\gamma$ attenuation factor
$e^{-\tau(\v',\mu')}$. The observed "time-averaged" flux is, then,
integration over the sphere,
\begin{eqnarray}
  F_{\varepsilon}&\propto&\int
  dS\f{1}{\Gamma^2(1-\beta_\Gamma\mu)^2}\f{d^3P^\prime}{d\Omega^\prime
  d\varepsilon^\prime dS}e^{-\tau(\v',\mu')}.
\end{eqnarray}
Assume isotropic emission power in the comoving frame, constant
emissivity along the sphere, and power law dependent on photon
energy, then
\begin{equation}
 \f{d^3P^\prime}{d\Omega^\prime
  d\varepsilon^\prime dS}\propto\varepsilon^{\prime-h+1}.
\end{equation}
Using $dS=2\pi R^2d\mu$ and eqs.
(\ref{eq:rdelay}),(\ref{eq:angletransfer}) and (\ref{eq:doppler}),
we have
\begin{equation}
  F_{\varepsilon}\propto
  t_d^2\Gamma^{-h+3}\varepsilon^{-h+1}f(\varepsilon;\Gamma),
\end{equation}
where
\begin{eqnarray}
   f(\varepsilon;\Gamma)=\int
  d\mu(1-\beta_\Gamma\mu)^{-h-1}e^{-\tau(\varepsilon',\mu')}.
\end{eqnarray}
Note that $f(\varepsilon;\Gamma)$ is the suppression factor of the
primary spectrum.

\begin{figure}[h]
\includegraphics[width=\columnwidth]{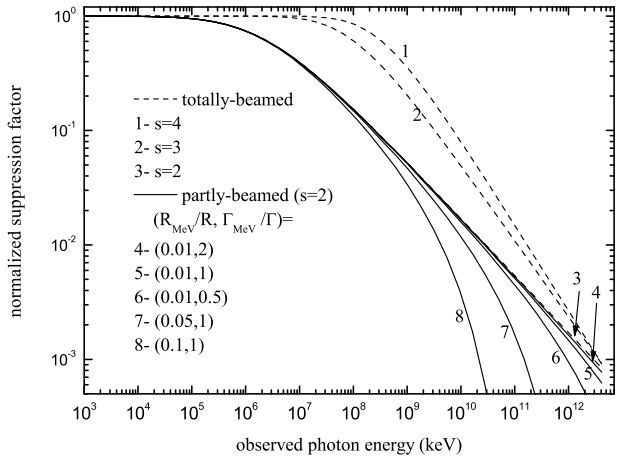}
\caption{In the two-zone case the suppression factor $f$, due to
attenuation by inner-originated and beamed target photons, as
function of the observed photon energy $\varepsilon$. Here the
target photons are assumed to be a single power law distribution.
All the lines correspond to the same parameters except those
labelled in the plot. The dashed lines correspond to the totally
beamed cases, calculated by using eq
(\ref{eq:tau_2zone_part_beam}), while the solid lines are for the
partly beamed cases using eq
(\ref{eq:tau_2zone}).}\label{fig:2zone}
\end{figure}

It is useful to analyze this factor analytically for the totally
beamed case. As shown in Appendix, for the single power-law target
photon distribution, $N(\e)=N_0\e^{-s}$ and using approximation
$\sigma(E)\approx\sigma_0E^{-2}$, the $f$ factor can be
approximated as
\begin{equation}\label{eq:ffactor}
f(\varepsilon;\Gamma)=\left\{\begin{array}{ll}
 1 &  \varepsilon<\varepsilon_{\rm br}\\
\left(\f{\v}{\v_{\rm br}}\right)^{\frac1{\bt}-1} &
\varepsilon>\varepsilon_{\rm br}\end{array} \right.,
\end{equation}
with the break energy at
\begin{equation}
  \v_{\rm{br}}=\frac{2}{\bt(1+z)^2}\left[\frac{(1+\bt)(m_e^2c^4)^{\bt}c^2t_d}
  {N_0\sigma_0d_L^2}\right]^{\frac1{\bt-1}}\Gamma^{\frac{2(1+\bt)}{\bt-1}}.
\end{equation}
We carry numerical calculation of $f$ factor, and show the result
in Fig \ref{fig:2zone}. The analytical result is a good
approximation.

Thus in the totally beamed case the spectrum is not affected until
$\v>\varepsilon_{\rm br}$, where the spectrum steepens by a factor
of $\frac1{\bt}-1$. Thus unlike in the case of isotropic target
photons, the spectrum is not cut off exponentially but show a
steepening power law. This can be easily understood- in the beam
target photon case, the HE photons always can escape if they
travel with a small enough angle with respect to the target beam.

The break energy is LF-dependent, thus the detection of the break
in the spectrum can be used to measure the LF of GRBs. For GRB
090510, the break energy is $\varepsilon_{\rm br}\approx1$ GeV for
$\Gamma=600$ and $t_d=0.1$s.

In Fig \ref{fig:2zone} we also show the numerical results for
partly beamed cases. As can be seen, for the energy range of
interests, say, $<1$~TeV, the partly beamed cases with
$\Gamma_{\rm MeV}=\Gamma$ approaches the totally beamed case when
$R_{\rm MeV}/R<0.1$, which is just the case we are considering
because $\delta t\ll t_d$. We also illustrate the small effect of
LF variation by showing the cases of $\Gamma_{\rm
MeV}/\Gamma=0.5-2$ with $R_{\rm MeV}/R=0.01$. Indeed when
$\Gamma_{\rm MeV}>\Gamma$ as expected, the MeV photons are more
strongly beamed then the situation is more approaching the totally
beamed case. Even when $\Gamma_{\rm MeV}/\Gamma\la1$ there is only
very little effect at very high energy (see Fig \ref{fig:2zone})
since $R_{\rm MeV}/R\ll1$ and the MeV photons are still highly
beamed. Thus we will ignore the effect of variation of LFs and
only consider $\Gamma_{MeV}=\Gamma$ in the following calculations.

It should be noted here that in the two-zone case, besides the HE
absorption due to the inner-coming beamed MeV photons, the
absorption due to interactions with photons locally originated
from the HE emission region can also contribute to the total
optical depth. This adds an extra attenuation factor
$e^{-\tau_{\rm self}(\varepsilon)}$ in the resulted spectrum,
where the optical depth $\tau_{\rm self}(\varepsilon)$ is given by
eq (\ref{eq:tau}) with $R$ being the HE emission region radius eq
(\ref{eq:rdelay}) instead. We also consider this absorption in the
following case studies.

\section{Case studies}
In this section we study the three bright Fermi-LAT GRBs 080916C,
090510 and 090902B, and constrain their LFs with assumptions of
one-zone or two-zone origins.

\subsection{GRB 080916C}
This is a bright long GRB, with a duration of $\sim50$s and 145
photons detected above 100 MeV, among which 15 are beyond 1 GeV
and 1 beyond 10 GeV. The redshift is quite high, $z=4.35$, so that
the isotropic-equivalent energy is turned out to be
$E_{iso}=8.8\times10^{54}$ erg, the largest energy measured so far
\citep{a09a}.

The wide energy range spectrum of this GRB is well fit by a single
Band function, which may imply that all radiation is originated
from one region. Indeed with one-zone assumption, the
one-component spectrum favors synchrotron origin over IC emission,
and the spectral slopes can be understood in the frame work of
synchrotron emission model \citep{wang09}. Using the time interval
3.58-7.68s, and under one-zone assumption, the LF has been
constrained to be $\Gamma>900$ by \cite{a09a}. It should be noted
that the constraint is variability time dependent. In this
constraint $\delta t=2$s is adopted from GBM light curve. However
INTEGRAL also detected this GRB and show variability time in MeV
range much shorter, $\delta t<100$~ms. With this shorter
variability time, we constraint the LF to satisfy $\Gamma>1130$
(Fig \ref{fig:1zone}).

However, the onset of $>100$~MeV emission is $\sim4$~s delayed
relative to MeV emission; and in time bin "b" the bulk emission
shifts toward later time as the photon energy increases, as
pointed by \citep{a09a}, and the shift is $1$-s scale, much longer
than the variability times in MeV light curves, $\delta t<100$~ms,
as noted by \cite{li10}. These temporal behaviors suggest that HE
emission may have different origins and larger emission regions
than the MeV one. Indeed, the "single" spectral component favors
synchrotron over inverse Compton radiation, however, as pointed by
\cite{li10}, the observed highest energy photon in this GRB cannot
be generated by synchrotron radiation, implying different
component/origin for the HE emission.

Here we consider a simple two-zone case, where the ejecta that
produce HE emission $>\varepsilon_0$ is released with a delay
$t_d=1-4$s relative to that produce MeV emission. It is hard to
determine the threshold energy $\varepsilon_0$ currently, but we
take $\varepsilon_0\ga30$~MeV, due to the different temporal
behaviors above $30$~MeV. We use the observed flux and spectrum to
calculate the optical depth due to absorption by inner-coming,
beamed photons, and only use that at $>\varepsilon_0$ to calculate
the optical depth due to self absorption by local-originated
photons from the HE emission region. With the sum of these two
optical depths we can calculate the suppression $f$ factor to
modify the original HE emission that is free of absorption. We
consider the time interval 3.58-7.68s following \cite{a09a}, and
assume the observed HE spectrum as the original one.

\begin{figure}[h]
\includegraphics[width=\columnwidth]{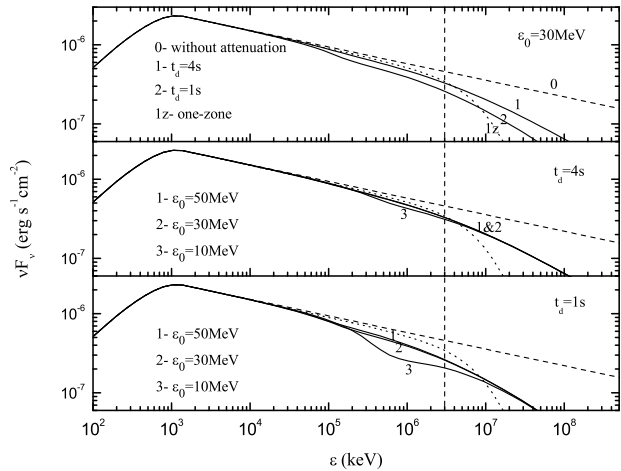}
\caption{The HE suppression in the two-zone case of GRB 080916C.
Here the observed, best-fit spectrum (see Table \ref{tab1}) is
assumed to be the original spectrum without attenuation (dashed
lines). The vertical line marks the observed highest photon energy.
The calculated attenuations take into account the absorptions by
both the inner- and local-originated photons. The upper panel shows
the case of fixed threshold energy $\varepsilon_0=30$~MeV, while the
middle and bottom panels show the attenuation varies with a range of
threshold energy $\varepsilon_0=$10-50~MeV for delayed time of
$t_d=4$ and 1~s respectively. In all these two zone cases
$\Gamma=600$ is taken. Note that the apparent dips in the last two
panels show the contribution of self absorption. Also shown for
comparison is the exponential cutoff (dot lines) in the one-zone
case with $\Gamma=900$ and $\delta t=2$~s \citep{a09a}.
}\label{fig:916c}
\end{figure}

The result is presented in Fig \ref{fig:916c}. It can be seen
that, with $\Gamma=600$ and $\varepsilon_0=30$~MeV, the "self"
absorption is less important than the "beamed" absorption, and the
attenuated spectrum does not show sharp cutoff but a slight
steepening, as in Fig \ref{fig:2zone}, in contrast with the
one-zone case. We also show that taking $\epsilon_0\ga30$~MeV does
not change the conclusion much. Indeed, for a photon of 3~GeV, the
highest observed energy in the relevant time bin, and given
$\Gamma=600$ the threshold energy of $\gamma\gamma$ interaction is
$\epsilon_{\rm th}=\Gamma^2(m_ec^2)^2/3{\rm
GeV}(1+z)^2\approx1$~MeV, much smaller than 30~MeV. Moreover, we
try different $\Gamma$ values and find the break energy, where the
steepening happens, increases with $\Gamma$. $\Gamma\sim600$ can
be consistent with the observed spectrum in the two-zone case.
Finally, it should be noted that the self-absorption becomes
important when taking smaller threshold energy and time delay,
i.e., $\varepsilon_0=10$~MeV and $t_d=1$~s (for $\Gamma=600$).

\subsection{GRB 090510}
This is a short GRB with a duration of 2.1s, but very bright, with
18 photons at $>1$~GeV detected. Given the redshift
$z=0.903\pm0.003$ and the total (0.5-1.0s) energy fluence in the
10keV-30GeV band,
$(5.02\pm0.26)\times10^{-5}\textrm{erg}~\textrm{cm}^{-2}$, the
total isotropic-equivalent energy release is
$(1.08\pm0.06)\times10^{53}$ erg \citep{a09b}.

Using the spectrum in time interval 0.8 s-0.9 s which includes a
highest energy photon of 31 GeV and can be fitted by the Band
function plus a power-law component, the LF constraint in one-zone
case is $\Gamma>1200$ \citep{a09b}. Under one-zone assumption we
find that $\Gamma>990$ (Fig \ref{fig:1zone}). The two results are
in broad consistence, though our result is a little less than that
of \cite{a09b}, which can be due to the different definition of
$R$. Our defined $R$ is larger by a factor of 2.

Moreover, there are some distinct features in this GRB: the
time-integrated spectrum in time interval 0.5-1.0s is best fit by
two spectral components, Band-function component at low energy
plus power-law component dominating HE emission; the emission
above 30MeV is delayed by $t_d=248$~ms than those below 1~MeV as
shown by the data analysis in \cite{a09b}. These suggest that HE
may have different origin and/or emission region. Therefore we
consider the simple two-zone assumption for this GRB again. Since
there are two components in the spectrum that may be consistent
with the two components in the temporal behavior, we use the Band
function component to calculate the optical depth due to MeV
photon beam and use the power-law component for calculation of
self absorption. Thus we obtain the $f$ factor to modify the
power-law component, assuming that the observed best fit spectrum
as the original one without $\gamma\gamma$ attenuation.

\begin{figure}[h]
\includegraphics[width=\columnwidth]{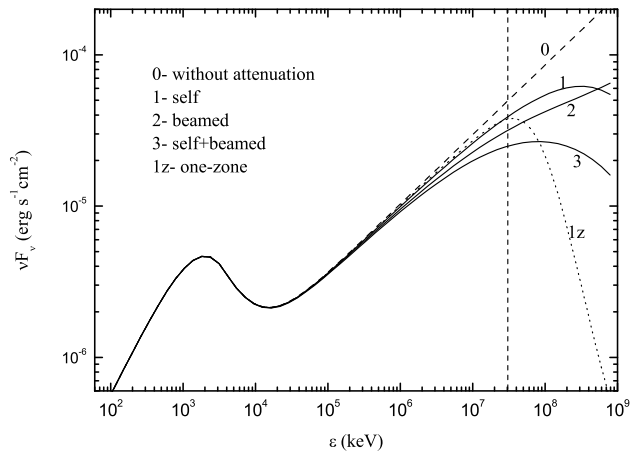}
 \caption{The HE suppression in the two-zone case of GRB
090510. $\Gamma=600$ is assumed, and $t_d=0.25$~s is taken due to
the data analysis by \cite{a09b}. Here "self" denotes the
absorption by local-originated photons in the HE emission region
itself, "beamed" by inner-originated and hence beamed photons, and
"beamed+self" by both inner- and local-originated photons. Also
shown for comparison is the one-zone case with $\Gamma=1200$ and
$\delta t=12$~ms \citep{a09b}.}\label{fig:510}
\end{figure}

The resulted spectra are shown in Fig \ref{fig:510}. We can find
that a LF of $\Gamma\sim600$ can be still consistent with the
observed spectrum. It should be noted that although the power law
component dominates in energy the Band function component still
dominates in photon number. So the absorption due to beamed MeV
photons can be more important. For the parameters taken, the self
absorption due to local originated photons contribute comparable,
though less important, effect, thus the attenuated spectrum is
steeper than beamed-MeV-photon only case, but still much smoother
than the sharp cutoff in one-zone case.

\subsection{GRB 090902B}
With the redshift of 1.822 this long, fairly strong GRB has an
isotropic-equivalent energy $E_{iso}=3.63\pm0.05\times10^{54}$ erg,
comparable with that of the highest-energy one GRB 080916C
\citep{a09c}. The duration in the energy interval 50-300 keV of
Fermi (GBM) is 22 s. The highest energy photon (33.4 GeV) in this
GRB is detected at 82 s after trigger, while that in the prompt
phase is 11.2 GeV and in interval of 9.6-13 s. Using this time
interval and one-zone assumption, \cite{a09c} constrain the LF to be
$\Gamma>1000$ \citep{a09c}, while we get, in Fig \ref{fig:1zone},
$\Gamma>830$.

\begin{figure}[h]
\includegraphics[width=\columnwidth]{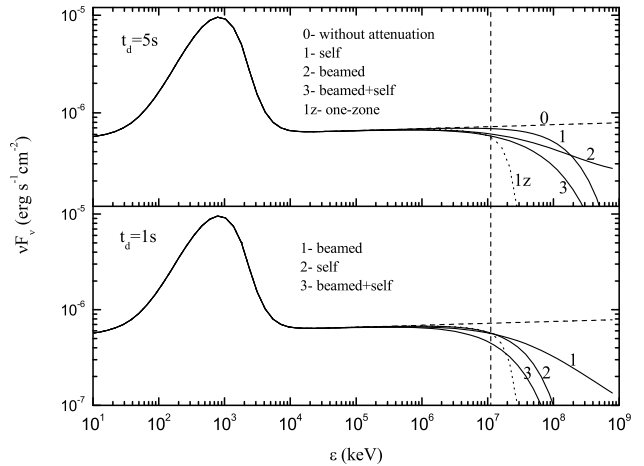}
\caption{The HE suppression in the two-zone case of GRB 090902B.
$\Gamma=600$ is assumed and the delayed time is taken to be
$t_d=5$~s (upper panel) and 1~s (lower panel). Also shown for
comparison is the one zone case with $\Gamma=1000$ and $\delta
t=53$~ms \citep{a09c}. }\label{fig:902b}
\end{figure}

Similar to GRB 090510, this GRB also has a distinct spectral
component fitted with a power law besides the Band function one. A
peculiar characteristic of its spectrum is that its power-law
component extends to lower band (<10keV). Similar to GRB 080916C,
there is an obvious delay of a few seconds in the HE onset. Look at
the time bin "b" in \cite{a09c}, the light curve peak seems also to
shift toward high energy, with a one-second delay. Then we consider
again a simple two-zone case taking $t_d=1-5$~s. Similar to GRB
090510, we use the Band function component as the beamed MeV target
photons of two-zone absorption, and the power-law component for the
self absorption at HE emission region. The total optical depth will
lead to $f$ factor calculation, which further modify the original
spectrum, assumed to be the observed spectrum. The results in Fig
\ref{fig:902b} suggest that $\Gamma\sim600$ can still be consistent
with observations.

\section{Discussion and conclusions}
We have revisited in this work the problem of constraining the GRB
LFs by the HE attenuation. Although this problem has been considered
by many previous works, two concerns that have been ignored in the
previous work have been emphasized here. First, we notice that in
the one-zone case in order to self-consistently calculate the
$\gamma\gamma$ optical depth one needs to consider the target
photons with HE spectral cutoff, other than extending to infinity.
This concern is important when the LFs are below a few hundreds, or
when the luminosity of GRBs are low. Second, we relax the one-zone
assumption and consider a simple two-zone case where the beaming of
target photons in the emission region should be taken into account.
Our results show that in the two-zone case, the $\gamma\gamma$
absorption does not lead to an abrupt spectral cutoff but a spectral
steepening. If the target photon energy distribution is with a power
law with photon index $\bt$ then the spectral slope is changed by a
factor of $\frac1{\bt}-1$. This also predicts that there should be
no spectral cutoff in the GRB spectra if the prompt emission is not
produced in one single region.

It should be noted that there are some attempts by other authors
to improve the approximation for the optical depth.
\cite{Baring06} concluded that the pair attenuation signature
appears as broken power-law rather than exponential cutoff by
considering the skin effect and introducing an attenuation
descriptor of $1/(1+\tau)$ instead of $e^{-\tau}$. \cite{Granot08}
considered the emission zone as a very thin layer producing
impulsive emission. They calculated in detail the opacity
evolution during a pulse, and claimed that the attenuation
signature can be different from that derived from the simple
one-zone approximation. Essentially, these two works still concern
one-zone problem, with $\Delta R\sim R_{\rm MeV}$. However in the
two-zone problem that we considered here, the HE and MeV emission
components are emitted at very different radii, with $\Delta R\gg
R_{\rm MeV}$, which leads to much smaller optical depth and hence
smaller LF at HE emission region.

Furthermore, we take our new concerns to analyze the spectra of the
three bright GRBs 080916C, 090510 and 090902B and found that in the
two-zone case a LF of $\Gamma\sim600$ can still be consistent with
the observed spectra. This relaxes the strict requirement,
$\Gamma>10^3$, in one-zone assumption.

We note that in the present observational situation where only
tens to hundreds HE photons detected in one GRB, a slight change
of the spectral slope is not easy to be identified. A single power
law may still fit the HE spectral tail.

We have considered a simple two-zone case here. However the
situation can be more complicated. The central engines of GRBs may
naturally create variabilities in a wide range of timescales, e.g.,
from $\sim1$~ms to $\sim10$~s. In the framework of internal shock
model, this will lead to kinetic dissipation in a wide range of
radii. Even in the single-timescale case, the internal collisions
will happen as the ejecta expand until the material is distributed
with velocity increasing with radius. In such case we will expect
multi-zone other than simple two-zone case. The time-integrated
spectrum-- note that the time interval with high enough photon
statistic is usually much larger than the variability time-- will be
contributed by the multiple regions. We also calculate cases with
$\Gamma<600$, the sum of the flux at the HE end can be comparable to
the original flux. This means that the spectra can be consistent
with a multi-zone case with the LF $\Gamma<600$.

The formation and acceleration of relativistic collimated GRB jets
are open questions. In the standard "fireball" model, the thermal
pressure can only accelerate the gas up to a LF $\Gamma\la10^3$
\citep[see, e.g.,][]{piran99,li10}. On the other hand, simulations
of magnetic-driven jets \citep[e.g.,][]{mhdjet} can generate jets
with the product of the LF and jet opening angle being
$\Gamma\theta_j\approx10-30$, which is consistent with pre-Fermi GRB
observations. However for the bright Fermi-LAT GRBs, \cite{cenko10}
constrained the jet opening angles by their afterglows, which,
combined with the large LF, $\Gamma>1000$ from $\gamma\gamma$
attenuation argument, suggests much larger values of
$\Gamma\theta_j$. We stress here that if relaxing the one-zone
assumption for GRB multi-band emission, LFs with "normal" values,
say, $\Gamma\la600$, can still be consistent with observations. This
relaxes further the theoretic problem of jet acceleration.

Recently, a similar paper, \cite{Zou10}, considering the same
two-zone absorption, is now in preprint. The main difference between
two papers is the rest frames for the optical depth calculation,
i.e., we consider the comoving frame of the HE emission region while
they consider the observer frame. They integrate over the region up
to $R_{\max}$, where the HE photon spatially leaves the MeV front,
while our integration corresponding to one dynamical time expansion
is equivalent to integration up to $2R$, where the generated HE
photon doubles its radius. Because both the number density of the
target photons and the angle between the travelling directions of HE
photon and the MeV front decrease rapidly with radius, the
interaction is strongly dominated by those at small radius, and
hence the upper limit of the integration is unimportant-- no matter
the upper limit is $R_{\max}(\gg R)$ or $2R$ the result is
practically the same. Furthermore, they use an "averaged" optical
depth to constrain the LF, which may not be appropriate since we
have shown that no sharp cutoff is expected in the two-zone case. We
consider more carefully the spectral profile due to suppression.
Finally, they neglect the self absorption in the local region which
may contribute significant effect as we show.

\begin{acknowledgements}
We thank Francesco de Palma for information. XHZ also thanks Z.G.
Dai and X.Y. Wang for helpful discuss. This work was partly
supported by the National Natural Science Foundation of China
through grant 10843007 and the Foundation for the Authors of
National Excellent Doctoral Dissertations of China.
\end{acknowledgements}

\begin{appendix}
\section{Derivation of the suppression factor for the totally beamed case}
Here we derive the suppression factor $f$ (eq.\ref{eq:ffactor}) in
the two-zone case, assuming the target photon distribution as a
single power law with $N(\e)=N_0\e^{-\bt}$. In the comoving frame of
the HE emission region, the target photon distribution can be given
by eq (\ref{eq:densitydistr}).

The cross section of $\gamma\gamma$ collisions in the relativistic
limit ($E\gg m_ec^2$) is $\sigma(E)\approx \sigma_0E^{-2}$, where
$\sigma_0=(3/8)\sigma_T(m_ec^2)^2[2\ln(2E/m_ec^2)-1]$ weakly depend
on $E$ and can be considered as constant due to roughly a constant
of $2\ln(2E/m_ec^2)-1$, which is as an approximation taken as 3.
With these approximations the $\gamma\gamma$ optical depth
(eq.\ref{eq:tau_2zone}) can be reduced to
 \bea
\tau=C_1\Gamma^{-\bt-3}
\varepsilon^{\prime\bt-1}(1-\mu^\prime)^{\bt}(1+z)^{\bt-1},
~~C_1=N_0\sigma_0(2m_e^2c^4)^{-\bt}d_L^2(\bt c^{2}t_d)^{-1}.
 \eea

Using the transformations of eqs (\ref{eq:angletransfer}) and
(\ref{eq:doppler}), the optical depth further becomes
 \bea
\tau= C_2\varepsilon^{\bt-1}(1-\mu)^{\bt}(1-\beta_\Gamma\mu)^{-1},
~~C_2=C_12^{\bt}(1+z)^{2(\bt-1)}\Gamma^{-4}.
 \eea

As $\theta$ increases ($\mu$ decreases) $\tau$ increases. Let us
define the minimum $\mu_{\min}$ where
$\tau(\varepsilon,\mu_{\min})=1$, then at $\mu<\mu_{\min}$ the
emission at $\v$ is significantly absorbed. Using the
approximation that $e^{-\tau}=1$ when $\tau<1$ and $e^{-\tau}=0$
when $\tau>1$, the expression for $f$ factor is approximated by
 \bea
f(\varepsilon;\Gamma)\approx&\int^{1}_{\mu_{\min}}d\mu(1-\beta_\Gamma\mu)^{-h-1}
\approx (1-\beta_\Gamma)^{-h-1}(1-\mu_{\min}).
 \eea
The second equality holds for $\mu_{\min}\rightarrow1$.

From the definition of $\mu_{\min}$, we can solve out
$\mu_{\min}$,
 \bea\label{eq:mu_min}
1-\mu_{\min}=C_2^{-\frac1{\bt}}(1-\beta_\Gamma\mu_{\min})^{\frac1{\bt}}\varepsilon^{\frac1{\bt}-1}\approx
C_2^{-\frac1{\bt}}(1-\beta_\Gamma)^{\frac1{\bt}}
\left[1+\f{\beta_\Gamma(1-\mu_{\min})}{\bt(1-\beta_\Gamma)}\right]\varepsilon^{\frac1{\bt}-1},
 \eea
where the second equality in the above equation, again, holds for
$\mu_{\min}\rightarrow1$.

We can see that approximately,
\begin{equation}
  1-\mu_{\min}\propto\left\{\begin{array}{ll}
 \varepsilon^{\frac1{\bt}-1} &~~
 1-\mu_{\min}<\frac{\bt}{\beta_\Gamma}(1-\beta_\Gamma)\\
\rm{const.} &~~
   1-\mu_{\min}>\frac{\bt}{\beta_\Gamma}(1-\beta_\Gamma)
\end{array}.
\right.
\end{equation}

The critical condition
\begin{equation}\label{eq:critical_condition}
  1-\mu_{\min}=\frac{\bt}{\beta_\Gamma}(1-\beta_\Gamma)
\end{equation}
corresponds to a "break energy" $\varepsilon_{\rm br}$.
Substituting eq (\ref{eq:critical_condition}) into the both sides
of the first equality in eq (\ref{eq:mu_min}), we get
\begin{equation}
  \v_{\rm{br}}=\frac{2}{\bt(1+z)^2}\left[\frac{(1+\bt)(m_e^2c^4)^{\bt}c^2t_d}{N_0\sigma_0d_L^2}\right]^{\frac1{\bt-1}}\Gamma^{\frac{2(1+\bt)}{\bt-1}}.
\end{equation}
$\v>\v_{\rm br}$ corresponds to
$1-\mu_{\min}<\frac{\bt}{\beta_\Gamma}(1-\beta_\Gamma)$ and
$\v<\v_{\rm br}$ to
$1-\mu_{\min}>\frac{\bt}{\beta_\Gamma}(1-\beta_\Gamma)$. So we can
write
 \bea
f(\v;\Gamma)\propto1-\mu_{\min}\propto\left\{\begin{array}{ll}
 \varepsilon^{\frac1{\bt}-1} &~~
 \varepsilon>\varepsilon_{\rm br}\\
\rm{const.} &~~
  \varepsilon<\varepsilon_{\rm br}
\end{array},
\right.
 \eea
or, after an arbitrary normalization,
\begin{equation}
f(\varepsilon;\Gamma)=\left\{\begin{array}{ll}
 1 &  \varepsilon<\varepsilon_{\rm br}\\
\left(\f{\v}{\v_{\rm br}}\right)^{\frac1{\bt}-1} &
\varepsilon>\varepsilon_{\rm br}\end{array}. \right.
\end{equation}
\end{appendix}

\end{document}